\documentclass[]{iopart}

\expandafter\let\csname equation*\endcsname\relax
\expandafter\let\csname endequation*\endcsname\relax
\usepackage[unicode]{hyperref}
\usepackage[all]{hypcap}
\usepackage{amssymb,amsmath,verbatim,mathtools,needspace,enumitem,etoolbox,graphicx,microtype}
\usepackage[usenames,dvipsnames]{xcolor}
\usepackage[T1]{fontenc}
\usepackage[utf8]{inputenc}
\usepackage{pifont}
\usepackage{yfonts}
\usepackage{url}
\usepackage[normalem]{ulem}
\definecolor{linkcolor}{rgb}{0.0,0.3,0.5}
\hypersetup{colorlinks=true,linkcolor=linkcolor,citecolor=linkcolor,filecolor=linkcolor,urlcolor=linkcolor}

\newcommand\orcid[1]{\href{https://orcid.org/#1}{$\!\!$\includegraphics[scale=0.006]{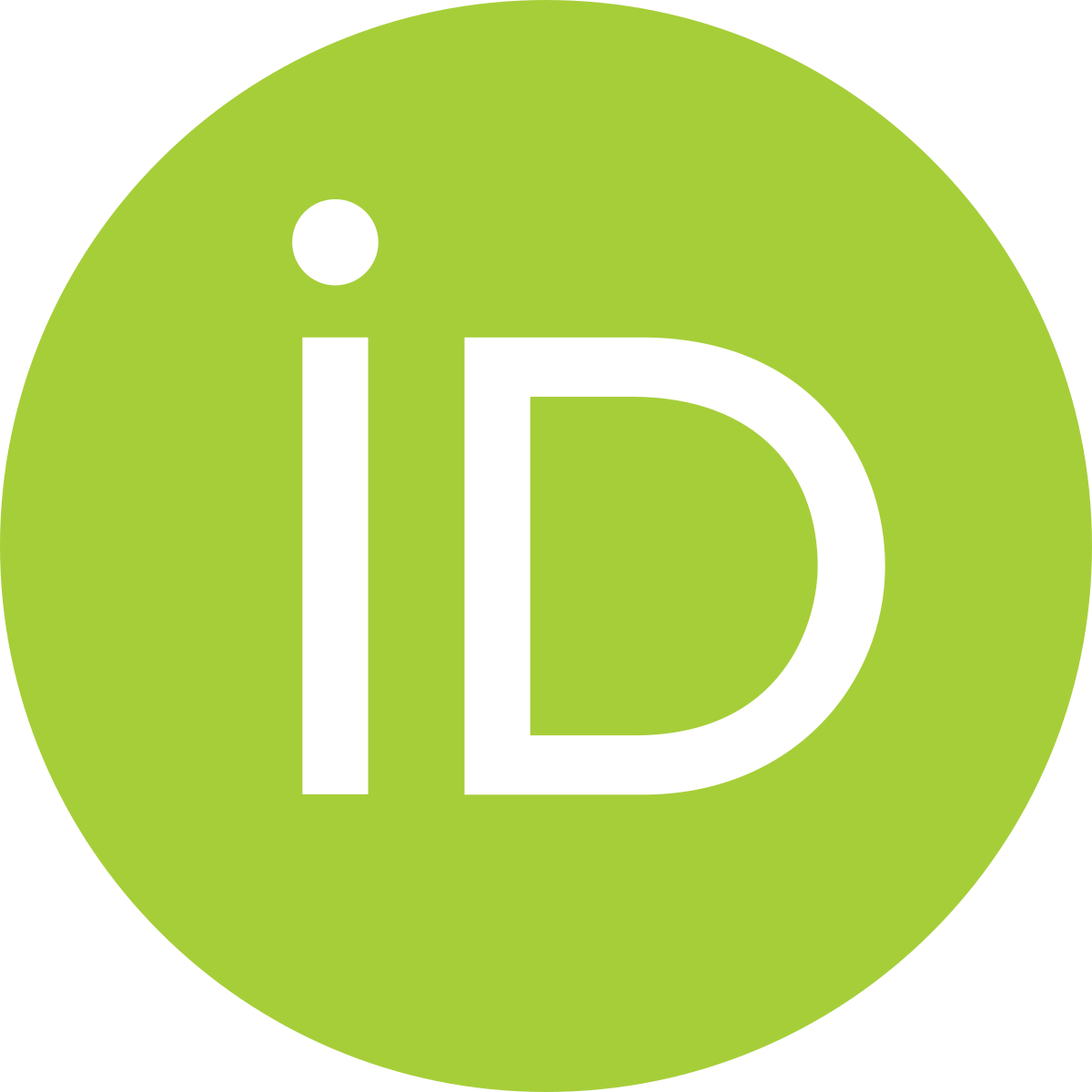} $\!\!$}}

\usepackage[numbers,compress,square]{natbib}

\newcommand{\ssim}{\mathchar"5218\relax\,} 

\newcommand\ddfrac[2]{\frac{\displaystyle #1}{\displaystyle #2}}

\graphicspath{{figs/}}

\newcommand{\bham}{{School of Physics and Astronomy \& Institute for Gravitational Wave Astronomy, University of Birmingham, Birmingham, B15 2TT, UK}}
\newcommand{\milan}{{Dipartimento di Fisica, Universit\`a degli Studi di Milano, Via Celoria 16, Milano, 20133, Italy}}
\newcommand{\caltech}{{TAPIR 350-17, California Institute of Technology, 1200 E California Boulevard, Pasadena, CA 91125, USA}}

\begin{document}

\begin{center}
\title[L.~Reali et al.]{Mapping the asymptotic inspiral of precessing \\binary black holes to their merger remnants}
\end{center}

\author{
Luca Reali$^{1,2,*}$ \orcid{0000-0002-8143-6767},
Matthew Mould$^{1}$ \orcid{0000-0001-5460-2910},
Davide Gerosa$^{1}$ \orcid{0000-0002-0933-3579},\\
Vijay Varma$^{3}$ \orcid{0000-0002-9994-1761}
}
\vspace{0.1cm}
\address{$^{1}$~\bham}
\address{$^{2}$~\milan}
\address{$^{3}$~\caltech}
\setcounter{footnote}{0}

\ead{\href{mailto:luca.reali@studenti.unimi.it}{luca.reali@studenti.unimi.it}}

\begin{abstract}
Multiple approaches are required to study the evolution of black-hole binaries. While the post-Newtonian approximation is sufficient to describe the early inspiral (even from infinitely large orbital separation), only numerical relativity can capture the full complexity of the dynamics near merger. We combine multi-timescale post-Newtonian integrations with numerical-relativity surrogate models, thus mapping the entire history of the binary from its asymptotic configuration at past-time infinity to the post-merger remnant. This approach naturally allows us to assess the impact of the precessional and orbital phase on the properties ---mass, spin, and kick--- of the merger remnant. These phases introduce a fundamental uncertainty when connecting the two extrema of the binary evolution.
\end{abstract}

\section{Introduction}
\label{sec:introduction}

Binary black holes (BHs) are prominent sources of gravitational waves (GWs).
Ten~\cite{2019PhRvX...9c1040A} or more \cite{2020PhRvD.101h3030V,2020ApJ...891..123N} BH mergers have been detected during the first two LIGO/Virgo observing
runs. Data from dozens of other candidates~\cite{GraceDB-O3-Public}
resulting from the third run are currently being
analyzed. With more interferometers becoming operational
both on the ground~\cite{2017arXiv171004823K,2013IJMPD..2241010U} and in
space~\cite{2017arXiv170200786A}, the observed population of BH binaries is
expected to grow dramatically in the near future.

Some of the events observed so far show  clear evidence for at least one BH with nonzero spin~\cite{2016PhRvL.116x1103A,2019PhRvD.100j4015C,2020arXiv200408342T,2019PhRvD.100b3007Z}. Systems where both BHs are spinning represent the most general and complex scenario: the three angular momenta of the binary (i.e. the spins of the two BHs and the orbital angular momentum) are all coupled to each other and precess about the total angular momentum of the system~\cite{1994PhRvD..49.6274A}. Meanwhile, the emitted GWs  dissipate energy and angular momentum, causing the orbital separation to shrink~\cite{1963PhRv..131..435P} and ultimately driving the BHs to merger. GWs also carry linear momentum, which results in a recoil (or kick) imparted to the center of mass of the system~\cite{1973ApJ...183..657B}. The final Kerr BH left behind following a merger is fully characterized by its mass $M_{\rm f}$, spin $\chi_{\rm f}$, and proper velocity $v_{\rm f}$.

Measurements of relative orientations of the three angular momenta can provide a powerful tool to distinguish stellar-mass BH binaries originating from different astrophysical formation channels~\cite{2013PhRvD..87j4028G,2017MNRAS.471.2801S,2017CQGra..34cLT01V,2017PhRvD..96b3012T,2017Natur.548..426F,2018ApJ...854L...9F}. In particular, spin directions are believed to be strong indicators to discriminate whether the two BHs formed in isolation~\cite{2018PhRvD..98h4036G}, star clusters~\cite{2016ApJ...832L...2R}, triple systems~\cite{2018MNRAS.480L..58A}, or active galactic nuclei~\cite{2020MNRAS.494.1203M}. For the case of supermassive BH binaries, spin orientations provide constraints on phenomena like star scattering and disk accretion \cite{2008ApJ...684..822B,2014ApJ...794..104S,2015MNRAS.451.3941G}.

Precessing BH binaries evolve on three different timescales. Let us denote the binary total mass with $M$, the orbital separation with $r$, and use geometrical units $c=G=1$. The BHs orbit about each other with period $t_{\rm orb}/M\propto (r/M)^{3/2}$, the angular momenta precess on a timescale  $t_{\rm pre}/M\propto (r/M)^{5/2}$ \cite{1994PhRvD..49.6274A}, and GW radiation-reaction takes place on the timescale $t_{\rm RR}/M\propto (r/M)^4$ \cite{1963PhRv..131..435P}. In the early inspiral where the dynamics is well described by the post-Newtonian (PN) approximation, one has $r\gg M$ and therefore $t_{\rm orb}\ll t_{\rm pre}\ll t_{\rm RR}$. This timescale separation implies that BH binaries complete a very large number of orbits and precessional cycles before they merge. By the time they become detectable, BHs will have lost memory of the precessional and orbital phase with which they formed. Astrophysical formation models aimed at describing BH binaries from assumptions on their progenitors cannot predict the orbital and precessional phase with which GW sources enter the detector sensitivity band.

The same reasoning applies to the properties (final mass, spin and kick) of the merger remnant  which are known to depend significantly on both  the precessional \cite{2010PhRvD..81h4054K,2010ApJ...715.1006K} and the orbital dynamics \cite{2008PhRvD..77l4047B,2018PhRvD..97j4049G}. These quantities are set by two phases, describing orbit and precession, which are essentially random.

The timescale hierarchy of spinning binary BHs has been exploited to develop
orbit-~\cite{2004PhRvD..70l4020S,2008PhRvD..78d4021R} and
precession-averaged~\cite{2015PhRvL.114h1103K,2015PhRvD..92f4016G} PN
approaches. Together, these schemes are  capable of evolving sources from their
asymptotic configuration at  $r\to\infty$ ($f\to0$, where $f$ is the GW
frequency) down to the small separations where the PN approximation ceases to
be an accurate physical description. From those small separations, estimates of
the  remnant parameters can be quickly obtained by evaluating
\emph{surrogate models}
trained on numerical-relativity (NR)
simulations~\cite{2017PhRvD..96b4058B,2019PhRvL.122a1101V,2019PhRvR...1c3015V}.  These models have been shown to reach accuracies comparable
to those of the underlying NR data. In particular,
Ref.~\cite{2019PhRvR...1c3015V} recently presented a surrogate model for the spin
dynamics and remnant properties of generically precessing binary
BHs covering mass ratios from 1:1 to 1:4 and dimensionless spin magnitudes up to 0.8.

In this paper, we combine the multi-timescale approach of
Ref.~\cite{2015PhRvL.114h1103K,2015PhRvD..92f4016G} with the NR surrogate model
of Ref.~\cite{2019PhRvR...1c3015V}. We design  a procedure to put  together
precession-averaged, orbit-averaged, and NR-surrogate evolutions. This allows
us to connect the asymptotic parameter space of precessing BH binaries to the
properties of the merger remnant, thus mapping the entire general relativistic
two-body problem from past to future time infinity.
We highlight how, for a given astrophysical configuration, the predictions of
the remnant properties suffer from theoretical uncertainties due to the
sampling of the orbital and precessional phases.

The paper is organized as follows. In Sec.~\ref{sec:methods} we illustrate our procedure for numerically evolving BH binaries; in particular we describe the implementation of multiple layers of PN/surrogate techniques as well as the random sampling of the  phases. In Sec.~\ref{sec:results} we present our results. We first focus on individual sources and assess the impact of phase sampling on the merger remnant properties.
We then consider BH binary populations to determine the impact of spin precession on the statistical distributions of the remnant parameters. We draw our conclusions in Sec.~\ref{sec:conclusions}.

\section{Multiple layers of black-hole evolution}
\label{sec:methods}

Let us consider spinning BH binaries on quasi-circular orbits. Let $m_{1}$ and $m_{2}$ be the masses of the two BHs, $M=m_1+m_2$ the total mass, $q=m_2/m_1\leq 1$ the mass ratio, $\eta=m_1 m_2/ M^2$ the symmetric mass ratio, $S_{i}=m_{i}^{2}\chi_{i}$ ($i=1,2$) the BH spin magnitudes (with $0\le\chi_{i}\le1$), and $L= m_1 m_2 \sqrt{r/M}$ the magnitude of the Newtonian orbital angular momentum. We write vectors in boldface, denote the corresponding unit vectors by hats, and their magnitudes as e.g. $L=|\mathbf{L}|$. Let us also define the following angles describing the spin directions 
\begin{align}
\label{angledef}
\cos\theta_{1}=\mathbf{\hat S}_{1}\cdot \mathbf{\hat L}\,,
&\qquad \cos\theta_{2}=\mathbf{\hat S}_{2}\cdot \mathbf{\hat L}\,, \notag
\\
\cos\theta_{12}=\mathbf{\hat S}_{1}\cdot \mathbf{\hat S}_{2}\,,&\qquad \cos\Delta\Phi = \frac{\mathbf{\hat S_1} \times \mathbf{\hat L}}{|\mathbf{\hat S_1} \times  \mathbf{\hat L} |} \cdot 
\frac{\mathbf{\hat S_2} \times \mathbf{\hat L}}{|\mathbf{\hat S_2} \times \mathbf{\hat L} |}\,.
\end{align}

\subsection{Early inspiral}
\label{sec:early inspiral}

Much like the inequality $t_{\rm orb}\ll t_{\rm pre}$ can be exploited to
average over the orbital motion, the relation $t_{\rm pre}\ll t_{\rm RR}$
allows averaging over spin precession. The computational cost of orbit-averaged
integrations blows up as $r/M\to \infty$ where infinitely many precession
cycles need to be tracked. On the contrary, precession-averaged evolutions are
capable of modeling the binary inspiral from its asymptotic conditions at past
time infinity. The cost one pays for precession (orbital) averaging is the ability to track the precessional (orbital) phase. 

The relative orientations of the three angular momenta $\mathbf{L}$, $\mathbf{S_1}$, and $\mathbf{S_2}$ at a given separation $r$ (or equivalently $L$) are  fully specified by the following three parameters:
\begin{itemize}
    \item the projected effective spin\footnote{This same parameter is more often indicated as $\chi_{\rm eff}$ in the GW literature.} 
    \begin{equation} \label{eq:xi}
        \xi = M^{-2} [(1+q)\mathbf{S}_1 + (1+q^{-1})\mathbf{S}_2] \cdot \hat{\mathbf{L}}\,,
    \end{equation}
    which is conserved on both $t_{\rm RR}$ and $t_{\rm pre}$ at 2PN order~\cite{2001PhRvD..64l4013D,2008PhRvD..78d4021R}\,;
    \item the magnitude of the total angular momentum 
    \begin{equation}
    J=|\mathbf{L}+\mathbf{S}_1+\mathbf{S}_2|\,,
    \end{equation} 
    which is constant on $t_{\rm pre}$ but slowly varies on $t_{\rm RR}$\,;
    \item the magnitude of the total spin 
    \begin{equation}
    S=|\mathbf{S}_1+\mathbf{S}_2|\,,
    \end{equation}
    which oscillates between two extrema $S_{+}\equiv\max_{\xi,J}(S)$ and $S_{-}\equiv\min_{\xi,J}(S)$ on the shorter timescale $t_{\rm pre}$\,.
   \end{itemize} 

The BH binary inspiral can thus be modelled by averaging over  $S$ (the only parameter that varies on $t_{\rm pre}$) and solving a single ordinary differential equation for $J$ (the only other parameter that varies on on $t_{\rm RR}$). In practice, the parameter $S$ sets the phase of the coupled precessional motion of $\mathbf{L}$, $\mathbf{S_1}$, and $\mathbf{S_2}$.

Since $J\sim L \propto \sqrt{r}$ diverges at large separations, it is  convenient to introduce the auxiliary variable 
\begin{equation}\label{eq:kappadef}
\kappa \equiv \frac{J^2 - L^2}{2 L}\end{equation}
which instead asymptotes to a constant 
\begin{equation}\label{eq:kappaliim}
\kappa_\infty \equiv \lim_{r/M\to\infty}\! \kappa = \lim_{r/M\to\infty} (\mathbf{S_1}+\mathbf{S_2})\cdot \hat{\mathbf{L}}\,.
\end{equation}
The integration domain can be compactified by setting $u=1/(2L)$
such that initializing an evolution at $u=0$ corresponds to integrating over the entire past history of the binary. 

Performing a precession-averaged
integration at 1.5PN reduces to solving~\cite{2015PhRvD..92f4016G}  (see also \cite{2017PhRvD..95j4004C})
\begin{align}\label{eq:dkappadu}
\frac{d\kappa}{du} 
= \ddfrac{
\int_{S_-}^{S_+}   S^2 \left|\frac{dS}{dt}\right| ^{-1} dS}{\int_{S_-}^{S_+}  \left|\frac{dS}{dt}\right| ^{-1} dS}\,,
\end{align}
where $dS/dt$ is obtained directly from the 2PN orbit-averaged spin-precession equations, cf. Eq.~(26) in Ref.~\cite{2015PhRvD..92f4016G}.  
This scheme is implemented in the public code \textsc{precession}~\cite{2016PhRvD..93l4066G}.

A BH binary spin configuration at $r/M\to\infty$ can be equivalently described using either $\xi$ and $\kappa_\infty$, or the asymptotic projections of the two spins along the orbital angular momentum:
\begin{align} 
\cos\theta_{1\infty}&\equiv\lim_{r/M \to \infty}  \cos\theta_1= \frac{M^2\xi - \kappa_\infty(1+1/q) }{S_1(q-1/q)}\,,
\\
\cos\theta_{2\infty}&\equiv\lim_{r/M \to \infty}  \cos\theta_2 = \frac{\kappa_\infty(1+q)-M^2\xi}{S_2(q-1/q)}\,.
\end{align}

\subsection{Late inspiral}

As shown extensively in Ref.~\cite{2015PhRvD..92f4016G}, precession- and orbit-averaged PN techniques are in very good agreement for separations $r \gtrsim 50M$. In the late inspiral, the timescales $t_{\rm pre}$ and $t_{\rm RR}$ become more comparable and the precession-averaged approach loses accuracy. We transition from  precession- to  orbit-averaged integrations at an orbital separation $r_\mathrm{pre \to orb} = 1000 M$.
This value is well within the validity of the precession-averaged approach and allows keeping the computational cost of the orbit-averaged evolution under control. The robustness of this choice is investigated in Sec.~\ref{transthr}. 

Initializing an orbit-averaged evolution requires the mutual orientations of the spins and the orbital angular momentum, which can be specified by either $(\theta_1,\theta_2,\Delta\Phi)$ or equivalently $(\xi,J,S)$. Only two of these parameters are provided by the previous precession-averaged treatment which averages over $S$. 
We estimate a value $S_{\mathrm{pre} \to \mathrm{orb}}$ of $S$ at the transition threshold statistically by drawing a random sample from the probability distribution function
$p(S)\propto |dS/dt|^{-1}$ between $S_+$ and $S_-$. Intuitively, it is more (less)
likely to find the binary with a precessional phase $S$ if the ``velocity''
$dS/dt$ is smaller (larger).

Our orbit-averaged code employs corrections up to 2PN  in spin precession and 3.5PN (2PN) in radiation reaction for (non)spinning terms as reported in Eq.~(24-27) of Ref.~\cite{2016PhRvD..93l4066G}.

\subsection{Plunge and merger}

The mass, spin and recoil velocity of the merger remnant are estimated using
the NR surrogate models of Ref.~\cite{2019PhRvR...1c3015V}.
This includes the \textsc{NRSur7dq4} model for the spin dynamics, and the \textsc{NRSur7dq4Remnant} model for the remnant properties. Both models are trained on NR simulations with mass ratios $q\geq0.25$, spin magnitudes $\chi_1, \chi_2 \leq 0.8$, and generic spin directions. Surrogate remnant predictions outperforms  alternative fitting formulae by orders of magnitude~\cite{2019PhRvL.122a1101V,2019PhRvR...1c3015V} and have recently been shown to be critical to model and extract BH kicks from GW
signals~\cite{2018PhRvD..97j4049G,2020PhRvL.124j1104V}.

The surrogate fits are provided as a function of the initial
orbital frequency $\omega$, mass ratio $q$, and six Cartesian components of the
spins $\mathbf{S_1}$ and $\mathbf{S_2}$. The orbit averaged scheme used in the
previous step only captures the relative orientation of the spins and the
orbital plane, which corresponds to five spin degrees of freedom (i.e.
$\chi_1$, $\chi_2$, $\xi$, $J$, and $S$). The additional parameter needed is the
orbital phase $\varphi$, describing the location of the BHs on their orbits
with respect to the spins or, equivalently, the orientation of the in-plane
components of the spins with respect to the vector connecting the two BHs.

The PN orbit-averaged evolution is carried over until the orbital frequency reaches
$\omega_\mathrm{orb \to sur} = 0.025\,{\rm rad}/M$.
This falls well within the validity of the \textsc{NRSur7dq4} model, which only includes $\ssim 20$ orbits before the merger;\footnote{The code
of Ref.~\cite{2019PhRvR...1c3015V} also allows to wrap the surrogate evaluation
with a PN evolution to specify the spin directions at earlier times. That
functionality is not used in this paper; we only make use of the NR surrogate.}
tests are reported in Sec.~\ref{transthr}. The conversion between PN separation $r$  and orbital
frequency $\omega$ is performed using  the 2PN\footnote{This PN order is appropriate. The 2PN term in Eq.~(\ref{eq:rtof}) introduces variations in $\omega$ of about $1-5\%$ which are  subdominant, see Sec.~\ref{transthr}.} expression~\cite{1995PhRvD..52..821K}
\begin{align}\label{eq:rtof}
    \omega^2 ={}& \frac{M}{r^3} \Bigg\{1-(3-\eta) \frac{M}{r} -\sum_{i=1,2}\chi_i\cos\theta_i\bigg[2\left(\frac{m_i}{M}\right)^2+3\eta\bigg]\left(\frac{M}{r}\right)^{3/2}
    \notag \\
&+ \bigg[6+\frac{41}{4}\eta
+\eta^2-\frac{3}{2}\eta\chi_1\chi_2(\cos\theta_{12}-3\cos\theta_1\cos\theta_2)\bigg]\left(\frac{M}{r}\right)^2\Bigg\}\,.
\end{align}
At the transition frequency, we sample a
value $\varphi_{\mathrm{orb} \to \mathrm{sur}}$ of the orbital phase uniformly
in $[0,2\pi]$, since quasi-circularity implies binaries spend equal time at all phases. First, the \textsc{NRSur7dq4} spin dynamics is
    evolved from the initial orbital frequency $\omega_\mathrm{orb \to sur}$ to a time $t=-100M$
    before the waveform amplitude peak.  Second, the
\textsc{NRSur7dq4Remnant} model is evaluated to predict the remnant mass, spin,
and recoil velocity. 

The Gaussian-process regression algorithm~\cite{2006gpml.book.....R,
2019PhRvL.122a1101V} implemented in Ref.~\cite{2019PhRvR...1c3015V} returns
mean values $\overline{M}_\mathrm{f}$, $\overline{\chi}_\mathrm{f}$ and
$\overline{v}_\mathrm{f}$ and $1\text{-}\sigma$ errors $\sigma_{M_\mathrm{f}}$,
$\sigma_{\chi_\mathrm{f}}$ and $\sigma_{v_\mathrm{f}}$ of respectively the
remnant mass, spin, and kick. Unless otherwise specified, we estimate these
quantities by sampling values $M_\mathrm{f}$, $\chi_\mathrm{f}$ and
$v_\mathrm{f}$ from the corresponding normal distributions, thus
 incorporating the fitting errors of the NR surrogate model.

\subsection{Summary}

\begin{figure}[t]
    \centering
    \includegraphics[trim=0 0 0 13.4cm,clip, width=0.8\columnwidth]{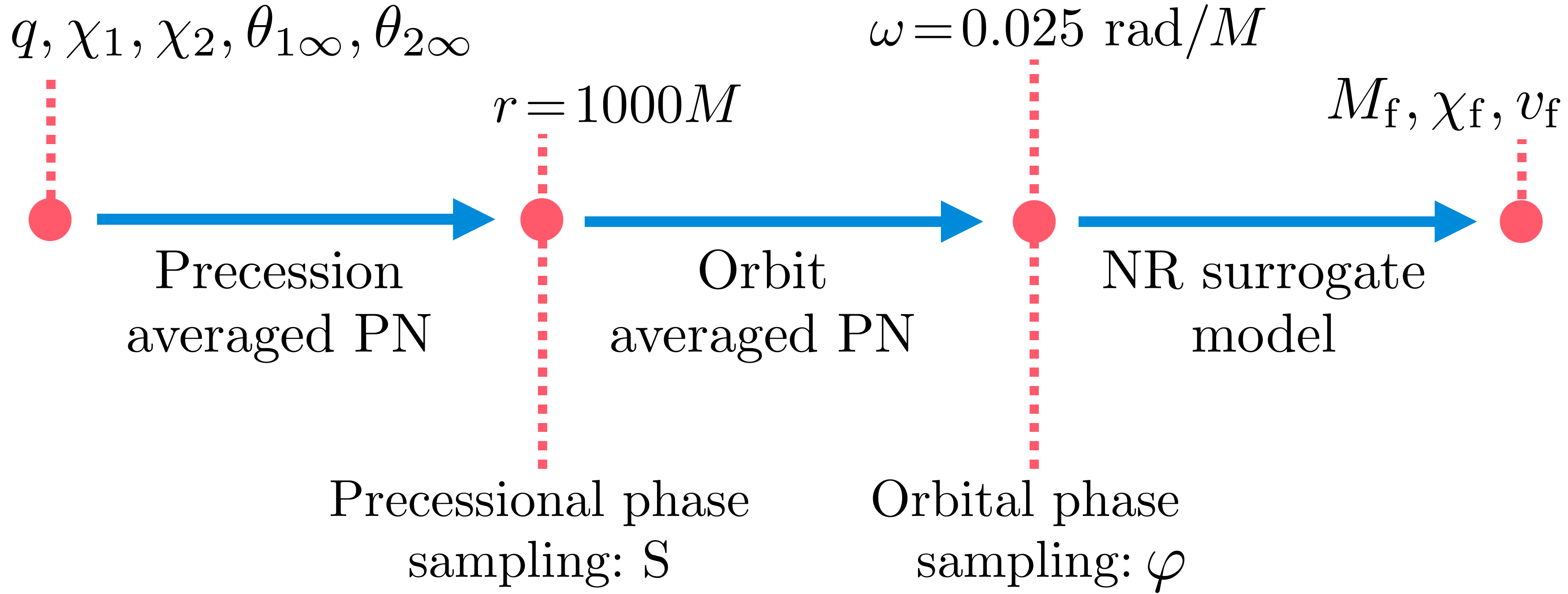}
    \caption{Schematic representation of our procedure. We initialize a precession-averaged PN evolutions at infinitely large separation, providing initial conditions on the mass ratio $q$, spin magnitudes $\chi_i$, and spin directions $\theta_{i \infty}$. At moderate separations $r_{\rm pre\to orb}=1000M$, we sample the precessional phase $S$ and switch to an orbit-averaged PN formulation. When the orbital frequency reaches $\omega_{\rm orb \to sur}=0.025\,{\rm rad}/M$, we sample the orbital phase $\varphi$ and evaluate the NR surrogate models to predict the remnant properties $M_{\rm f}$, $\chi_{\rm f}$, and $v_{\rm f}$.}
    \label{fig:procedure}
\end{figure}

Our procedure is summarized in Fig.~\ref{fig:procedure}. We map the entire
history of the binary from its asymptotic configuration at $r/M\to \infty$ to
its merger product. We start by initializing a precession-averaged PN evolution
at $u=0$ ($r=\infty$) for a binary with parameters $q$, $\chi_1$, $\chi_2$,
$\theta_{1\infty}$, and $\theta_{2\infty}$. We integrate
Eq.~(\ref{eq:dkappadu}) from the initial configuration to an orbital separation
$r_{\rm pre\to orb}=1000 M$. At this separation we sample the precessional phase $S$ according
to $p(S) \propto |dS/dt|^{-1}$ and integrate the PN orbit-averaged equations
until the orbital frequency reaches $\omega_{\rm orb \to sur}=0.025\,{\rm rad}/M$. We then sample
the orbital phase $\varphi$ uniformly from $[0, 2\pi]$ and evaluate the NR
surrogate model to obtain the parameters $M_{\rm f}$, $\chi_{\rm f}$, and $v_{\rm f}$.

Each point in the asymptotic space $(q,\chi_1,\chi_2,\theta_{1\infty},\theta_{2\infty})$ thus corresponds to a distribution in remnant space $(M_{\rm f},\chi_{\rm f}, v_{\rm f})$. In the following, we characterize the remnant properties using $90\%$ intervals and define 
\begin{equation}
\label{estimator}
\delta x = \frac{x_{95}-x_{5}}{2\, x_{50}}\,,
\end{equation}
where $x=\{M_{\rm f},\chi_{\rm f}, v_{\rm f}\}$ is any of the remnant properties and $x_n$ indicates the n-th percentile of its marginalized distribution.

\section{Impact of orbital and precessional phases}
\label{sec:results}

\subsection{Uncertainty of the black hole remnant}
\label{subsec:uncertainties}

\begin{figure}[t]
\centering
\includegraphics[width=0.8\columnwidth]{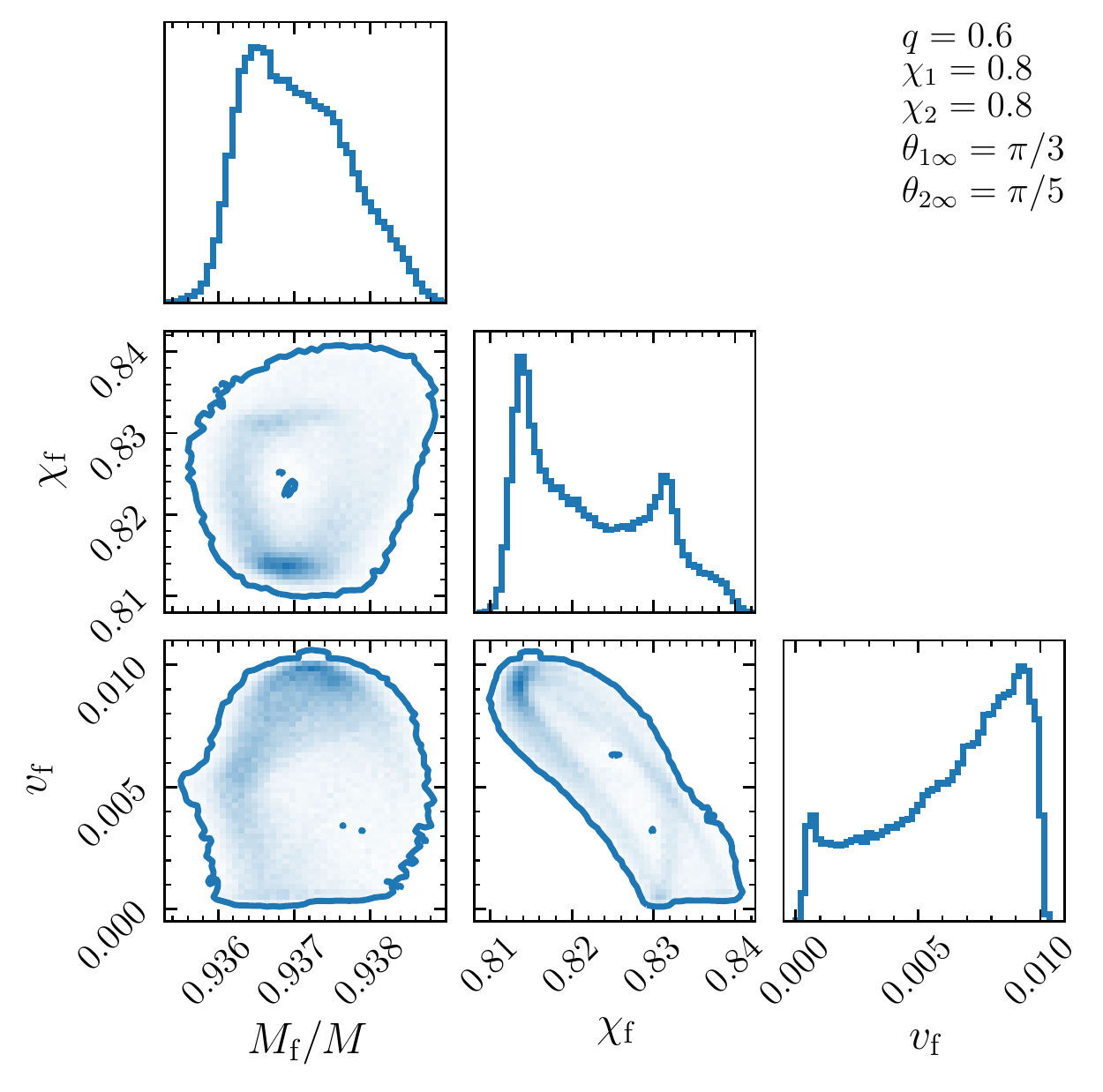}
\caption{Joint distribution of final mass $M_{\rm f}$, final spin $\chi_{\rm f}$ and final kick $v_{\rm f}$ for a single binary  with mass ratio $q=0.6$, spin magnitudes $\chi_1=\chi_2=0.8$ and asymptotic spin directions $\theta_{1\infty}=\pi/3$, $\theta_{2 \infty}=\pi/5$. Solid contours enclose the regions containing $99\%$ of the binaries.  This figure is produced with $10^5$ evolutions.}
\label{fig:cornerplot}
\end{figure}

We start by considering a single binary at $r/M \to \infty$ and studying the impact of the two phases. In the examples below we selected a source with $q=0.6$, $\chi_1=\chi_2=0.8$, $\theta_{1\infty}=\pi/3$, and $\theta_{2\infty}=\pi/5$ but the trends we highlight are generic. 

We first evolve this same binary $10^5$ times repeatedly sampling over all three sources of uncertainty: the precessional phase $S_\mathrm{pre \to orb}$, the orbital phase $\varphi_\mathrm{orb \to sur}$, and the surrogate Gaussian errors. Figure~\ref{fig:cornerplot} shows the resulting distribution of the BH remnant properties. GWs dissipate energy and angular momentum relatively gradually during the inspiral; on the other hand, the vast majority of the linear momentum is emitted during the last few orbits and plunge~\cite{2008PhRvD..77l4047B,2018PhRvD..97j4049G}. The exact configuration with which the binary approaches merger, therefore, plays a more crucial role for $v_{\rm f}$ compared to $M_{\rm f}$ and $\chi_{\rm f}$.
For this specific binary, the final mass presents a median value of $M_{\rm f} \simeq 0.937 M$ with a dispersion $\delta M_{\rm f} \simeq 0.1\%$. The final spin presents uncertainties that are up to an order of magnitude larger: the 90\% confidence interval extends for about $\delta \chi_{\rm f} \simeq 1.4\%$ from a median value $\chi_{\rm f} \simeq 0.82$. The kick is subject to variations which are even more extreme: possible values of the final recoil range from $v_{\rm f}\ssim 0$ to $\ssim 0.01$.

\subsection{Contributions to the remnant uncertainty}
\label{subsec:impact}

We now dissect the various contributions entering these uncertainties.

\begin{figure}[t]
\centering
\includegraphics[width=\textwidth, trim=1cm 0 0 1cm, clip]{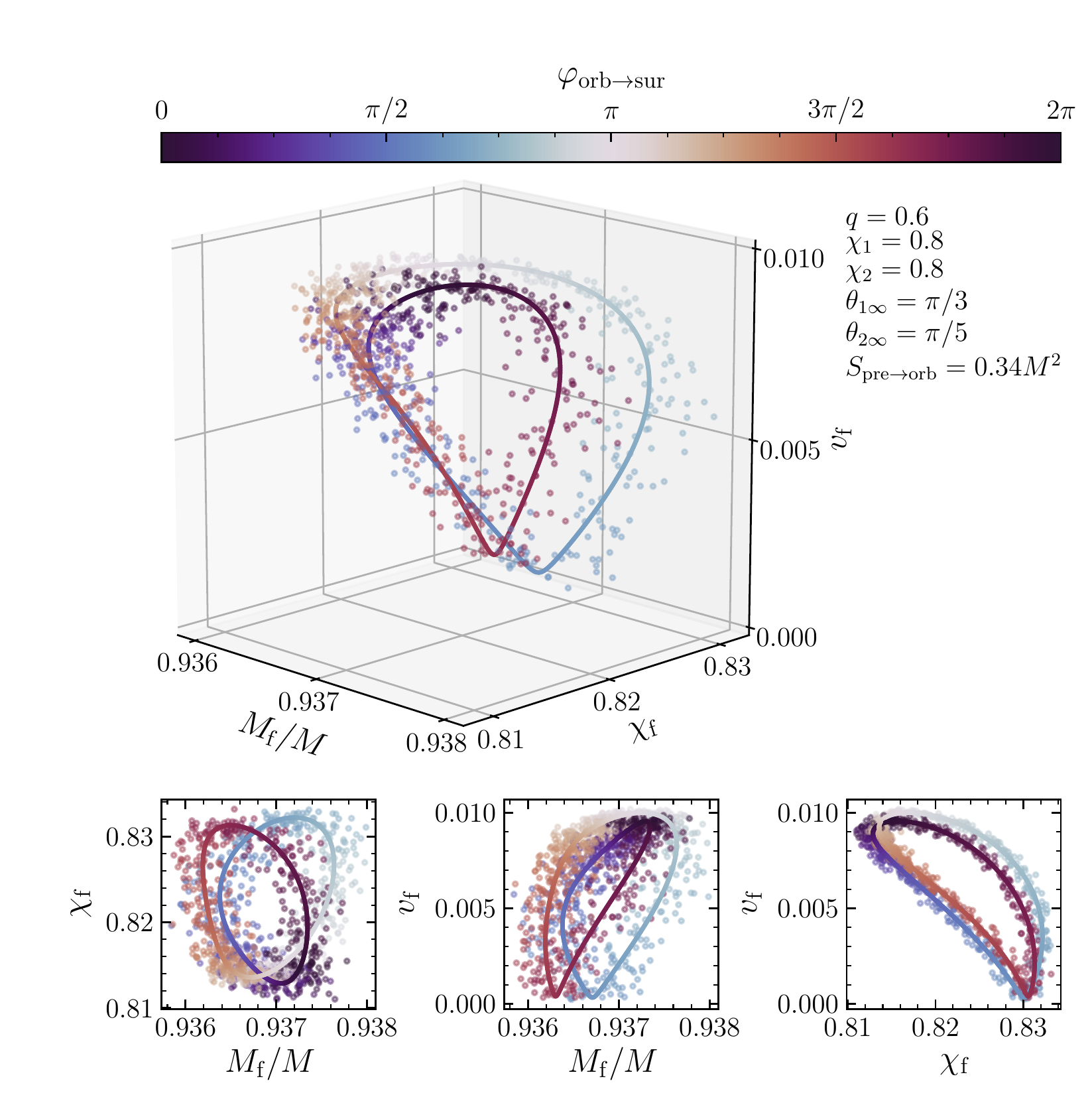}
\caption{The loop structure in the mass-spin-kick space. We consider a generic binary initialized with parameters $q=0.6$, $\chi_1=\chi_2=0.8$, $\theta_{1\infty}=\pi/3$ and $\theta_{2\infty}=\pi/5$. The precessional phase is set to $S_\mathrm{pre \to orb}=0.34 M^2$. The orbital phase $\varphi_\mathrm{orb \to sur}$ is varied from $0$ to $2\pi$ as indicated on the colour bar at the top. The mean values of the remnant properties returned by the NR surrogate are displayed with a solid curve; circles indicate a statistical sample extracted from the corresponding Gaussian distributions.}
\label{fig:singleloop}
\end{figure}

\begin{figure}[t]
	\centering
	\includegraphics[width=\textwidth]{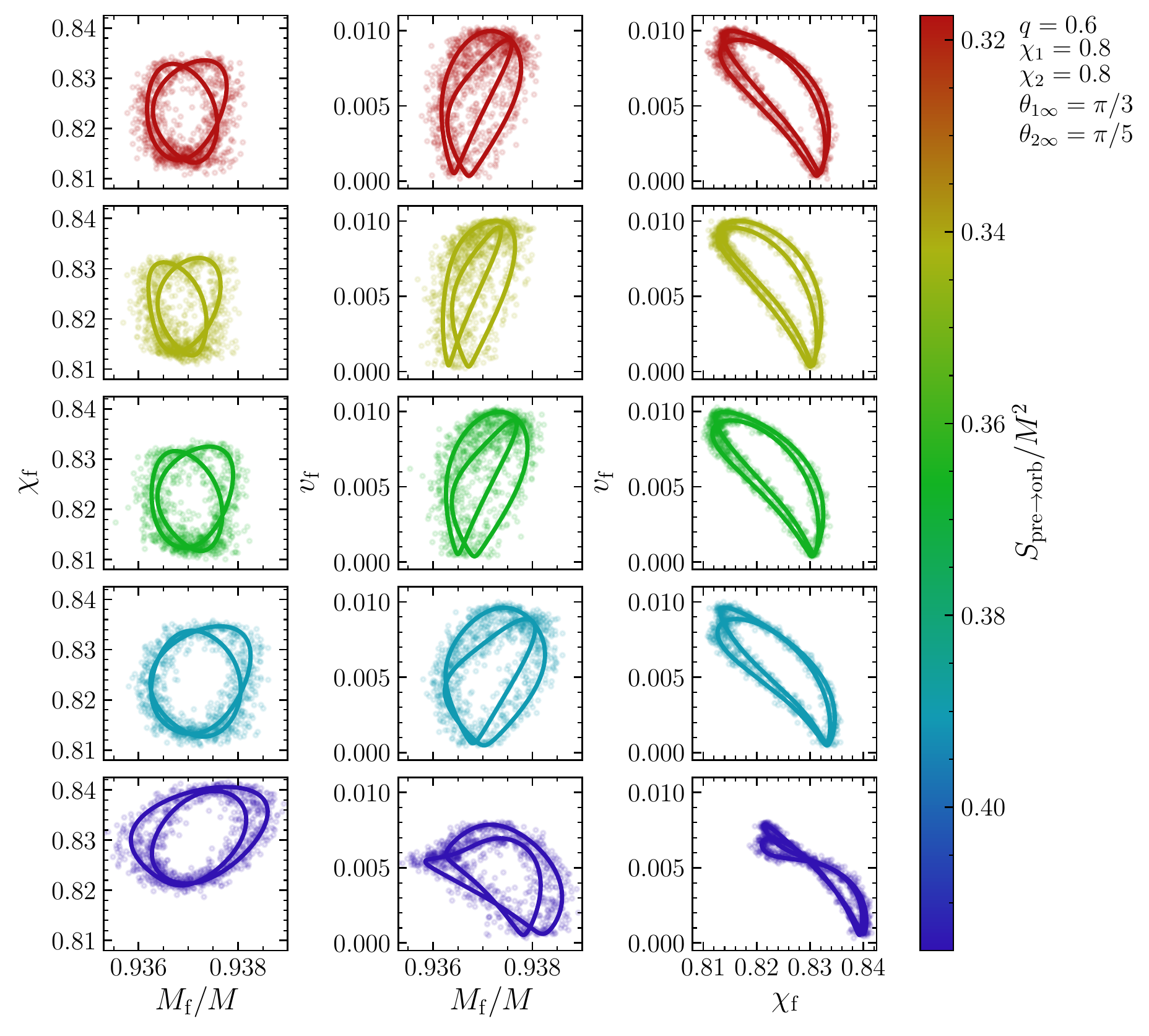}
	\caption{Final remnant properties as a function of the precessional phase $S_\mathrm{pre \to orb}$ measured at the transition between precession- and orbit-averaged PN evolution.  From left to right, we show the mass-spin, mass-kick and spin-kick planes. For this evolution we assume $q=0.6$, $\chi_1=\chi_2=0.8$, $\theta_{1\infty}=\pi/3$, and $\theta_{2\infty}=\pi/5$. Each row is generated similarly to Fig.~\ref{fig:singleloop}, by keeping $S_{\rm pre \to orb}$ fixed while varying the orbital phase $\varphi_\mathrm{orb \to sur}$ from $0$ to $2\pi$. The value of $S_{\rm pre \to orb}$ increases from top to bottom from $S_-$ to $S_+$ as reported on the color bar. In particular, the panels are produced with representative values $S_{\rm pre \to orb}\simeq 0.32,0.34, 0.37,0.39,0.41$ linearly spaced within the allowed range.}
	\label{fig:manyloops}
\end{figure}

Figure~\ref{fig:singleloop} shows the remnant properties of the same asymptotic configuration considered previously, but now assuming a fixed value of the  precessional phase $S_{\rm pre\to orb}=0.34 M^2$. We then vary $\varphi_{\rm orb\to sur}$ from $0$ to $2\pi$. The solid curves in Fig.~\ref{fig:singleloop} show the median values $\overline{M}_\mathrm{f}$, $\overline{\chi}_\mathrm{f}$ and $\overline{v}_\mathrm{f}$ returned by the NR surrogate; circles mark a sample of remnants extracted from the corresponding Gaussian distributions. 

The orbital-phase uncertainty on the remnant properties is due to the breaking down of the PN timescale hierarchy. The series of binaries depicted in Fig.~\ref{fig:singleloop} is computed for different phases $\varphi_\mathrm{orb \to sur}$ and fixed orbital frequency $\omega_{\rm orb \to sur}$. For $t_{\rm orb}/t_{\rm pre}\to 0$ and $t_{\rm orb}/t_{\rm RR}\to 0$, this sequence is equivalent to the same binary moved along a quasi-adiabatic orbit, and should just result in identical remnants. Deviations from these limits imply that, although these sources are identical up to $r_{\rm pre\to orb}$, they evolve into different configurations and thus result in different merger remnants.

By construction, the curve describing the surrogate median values is periodic
as $\varphi_\mathrm{orb \to sur}$ ranges from $0$ to $2\pi$. An approximate periodicity with period
equal to $\pi$ is also present, such that the curve  in
Fig.~\ref{fig:singleloop} appears as two, largely overlapping loops.  A period
equal to twice the orbital frequency corresponds to that of the dominant GW
emission. The separation between the two loops is typically smaller than the
Gaussian errors of the surrogate model, and thus largely disappears when these
are considered (scatter points).

For binaries with $q=1$ and $\mathbf{S_1}=\mathbf{S_2}$, a transformation $\varphi\to \varphi+\pi$ can be absorbed into a change of reference frame and thus results in identical values of $M_{\rm f}$, $\chi_{\rm f}$, and $v_{\rm f}$. The two-loop structure of Fig.~\ref{fig:singleloop} thus collapses into a single loop (in practice, this is only true within the errors of the surrogate model because this expected symmetry was not built in explicitly in Ref.~\cite{2019PhRvR...1c3015V}). For non-symmetric binaries, the $\pi$ periodicity is broken resulting in a more complex phenomenology.  In general, we find that for binaries with more extreme mass ratios $q\lesssim 0.25$, the loop of the median values becomes tighter and is eventually ``filled'' as the surrogate errors are considered.
 We refer to Refs.~\cite{2008PhRvL.100o1101B, 2008PhRvD..78b4017B} for a careful analysis of the expected symmetries. 

Figure~\ref{fig:manyloops} shows how the remnant properties vary with the  precessional phase $S_{\rm pre\to orb}$. While the precise shape of the curves depends on the sampled value, the general behavior we described remains valid: the remnant properties present an approximate period equal to that of the dominant GW mode, which is half  the orbital period. The region covered by the loops mildly changes with $S$, such that the weighted superpositions of all the loops in Fig.~\ref{fig:manyloops} returns the full distribution we first showed in Fig~\ref{fig:cornerplot}. Each individual contribution is weighted by the probability density function $p(S)\propto |dS/dt|^{-1}$, which is strongly peaked at the two extrema $S_-$ and $S_+$ \cite{2016PhRvD..93l4066G}. The loops with the larger and smaller values of $S_{\rm pre\to orb}$ (top and bottom row in Fig.~\ref{fig:manyloops}) provide the largest contributions to the overall distribution and can indeed be glimpsed when carefully inspecting  Fig.~\ref{fig:cornerplot}. The bimodality observed in the marginalized distributions on Fig.~\ref{fig:cornerplot} is also a direct consequence of this loop structure.

\subsection{Black-hole binary populations}
\label{subsec:populations}

We now consider BH binary populations and investigate the statistical properties of their remnants.

\begin{figure}[t]
\centering
\includegraphics[width=\textwidth]{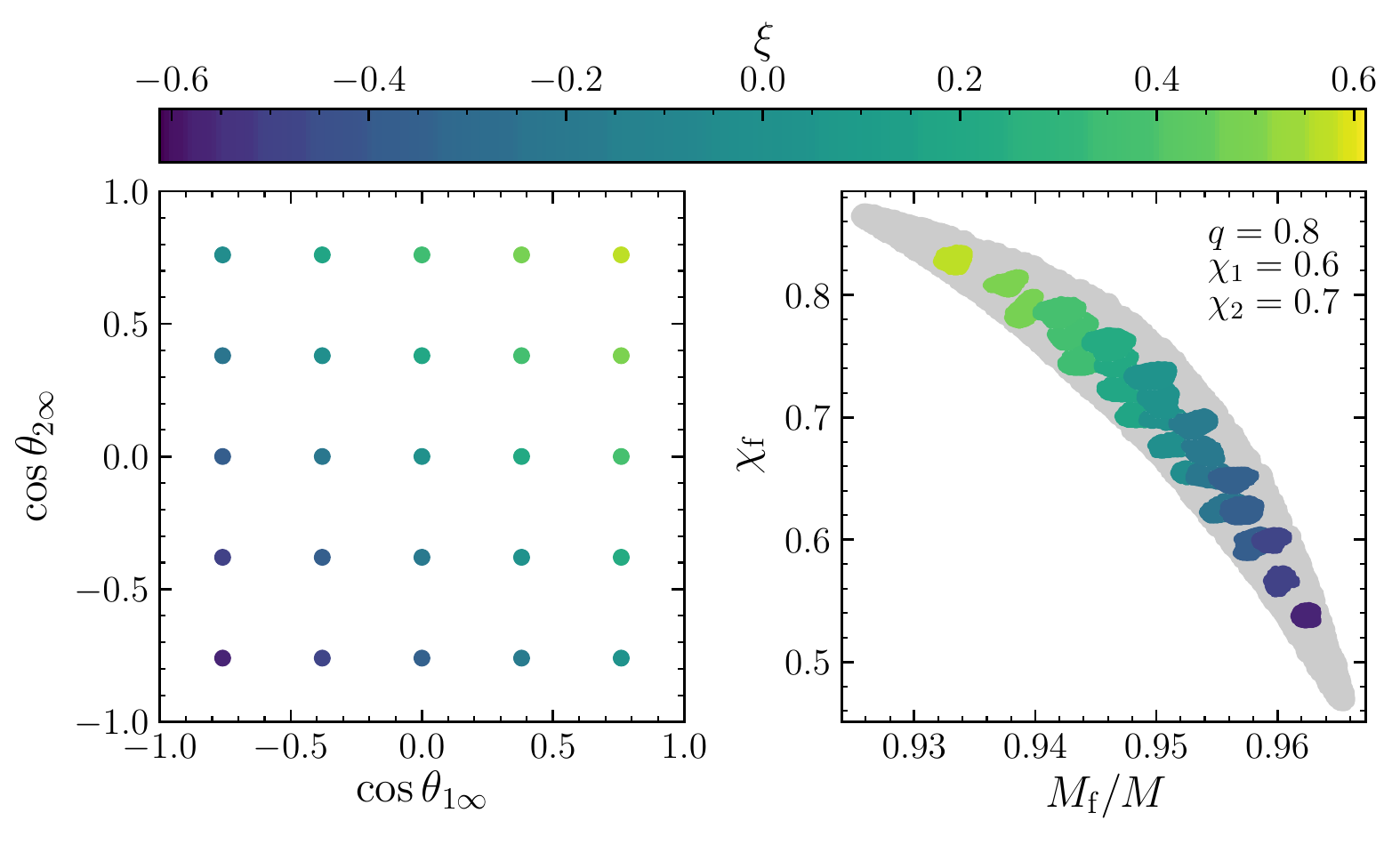}
\caption{Mapping of the $(\cos\theta_{\rm 1\infty},\cos\theta_{\rm 2\infty})$ plane (left panel) into the $(M_{\rm f},\chi_{\rm f})$ parameter space (right panel) for a binary BH population with fixed $q=0.8$, $\chi_{\rm 1}=0.6$ and $\chi_{\rm 2}=0.7$. The gray area on the right panel represents the whole accessible region of the $(M_{\rm f},\chi_{\rm f})$ plane for a population with these values of $q$, $\chi_1$, and $\chi_2$. Circles on the left panel mark a subsample of binaries with isotropically distributed spins at $r\to \infty$, colored according to the their projected effective spin $\xi$. Each of these configurations is mapped into a discrete region on the $(M_{\rm f},\chi_{\rm f})$ parameter space.}
\label{fig:clouds}
\end{figure}

In Fig.~\ref{fig:clouds}, we focus on a population with fixed $q=0.8$ and $\chi_1=0.6$ and $\chi_2=0.7$. First, we  map the entire $(\cos\theta_{\rm 1\infty},\cos\theta_{\rm 2\infty})$ plane into the $(M_{\rm f},\chi_{\rm f})$ parameter space (gray region in the right panel of Fig.~\ref{fig:clouds}). Final mass and final spin are strongly anti-correlated, with large values of $M_{\rm f}$ corresponding to small values of $\chi_{\rm f}$, and vice-versa. This is due to the orbital hang-up effect \cite{2001PhRvD..64l4013D, 2006PhRvD..74d1501C,2015CQGra..32j5009S}. Binaries with spins which are both aligned to the orbital angular momentum ($\theta_{1\infty}\simeq\theta_{2\infty}\simeq 0$, corresponding to larger positive values of $\xi$ in the upper right corner of the left panel of Fig.~\ref{fig:clouds}) inspiral for longer: they dissipate more energy via GWs and result in lighter remnant. Their angular momenta add constructively ($|\mathbf{L}+\mathbf{S_1}+\mathbf{S_2}|\simeq L+S_1+S_2$) resulting in a larger final spin. On the contrary, binary with spins which are largely anti-aligned plunge from afar but the spins and angular momentum partially cancel each other, thus producing remnants which are heavier and more slowly rotating. 

We then consider a subsample of 25 equispaced sources from the same population, shown as coloured dots in the left panel of Fig.~\ref{fig:clouds}. In particular, binaries are coloured according to the projected effective spin $\xi$ [Eq.~(\ref{eq:xi})]. We evolve each of these systems $10^4$ times, sampling over $S_\mathrm{pre \to orb}$, $\varphi_\mathrm{orb \to sur}$ and the surrogate Gaussian errors. The corresponding distributions in the remnant parameter space are shown in the right panel of Fig.~\ref{fig:clouds} with the same color scheme. The larger these regions, the more uncertain the remnant properties are. The size of the regions correlates with $\xi$: remnant uncertainties tend to be larger for $|\xi|\ssim 0$ and smaller for values at the extrema $|\xi|\ssim (\chi_1+q\chi_2)/(1+q)$.
This is because the former systems have larger precessional amplitudes $|S_+ - S_-|$ while for the latter the range where $S_\mathrm{pre \to orb}$ can be sampled is more limited \cite{2015PhRvD..92f4016G}.

\begin{figure}[t]
\centering
\includegraphics[width=\textwidth]{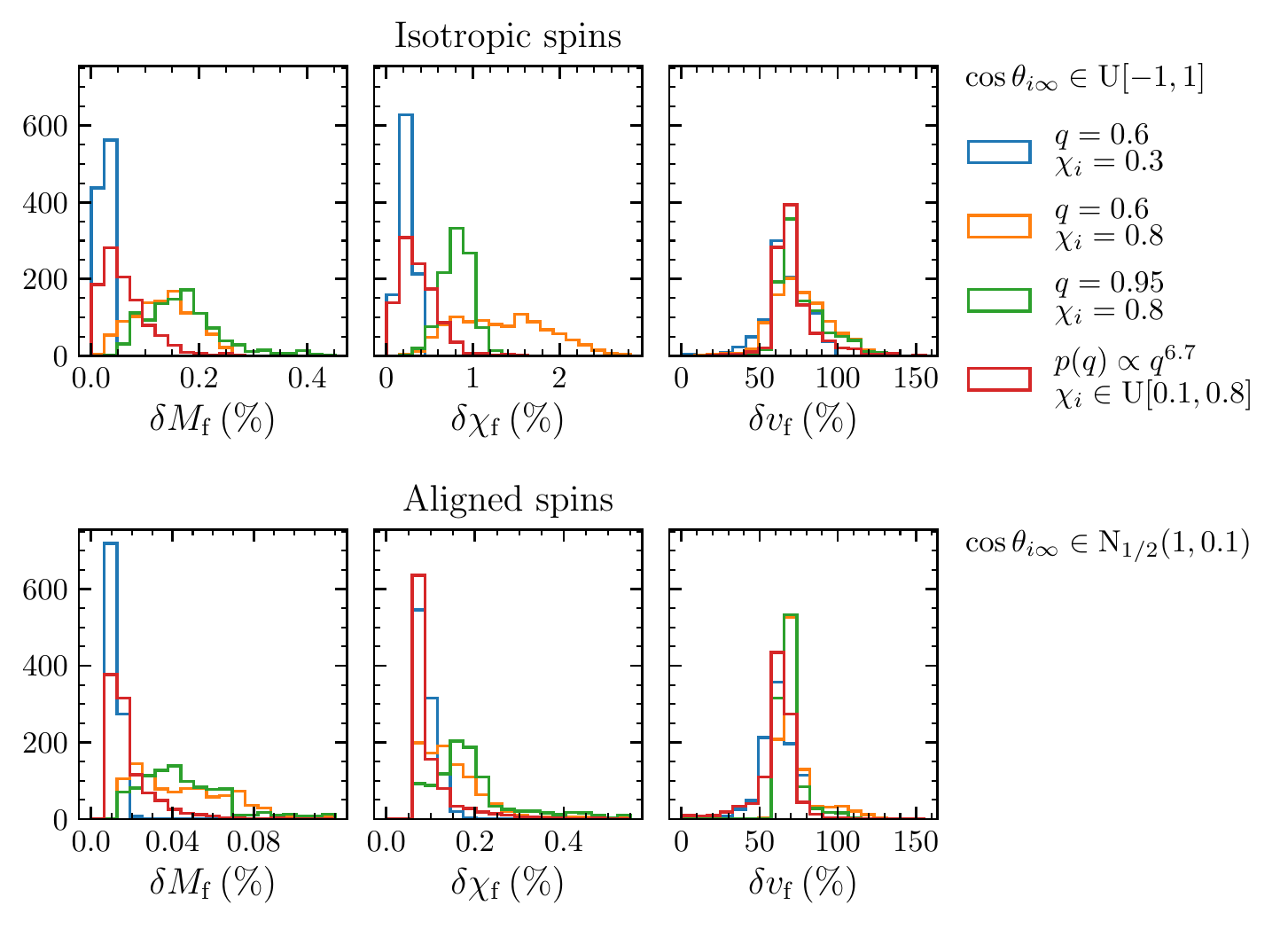}
\caption{Distributions of the relative uncertainties over final mass (left panel), spin (middle panel) and recoil velocity (right panel) of the remnant for different binary populations. The red histograms correspond to a sample of binaries with $\chi_{\rm 1}$, $\chi_{\rm2}$ uniformly distributed within the range $[0.1,0.8]$ and $q$ randomly extracted according to a distribution $p(q)\propto q^{6.7}$ in $[0.25,1]$ (see Ref.~\cite{2019ApJ...882L..24A}). The blue, orange and green histograms correspond to three binary samples with fixed mass ratios and spin magnitudes $q = 0.6$ and $\chi_1 = \chi_2 = 0.3$, $q = 0.6$ and $\chi_1 = \chi_2 = 0.8$, and $q = 0.95$ and $\chi_1 = \chi_2 = 0.8$, respectively. The populations shown in the top (bottom) panels have isotropic (preferentially aligned) spins at past time infinity.}
\label{fig:range}
\end{figure}

Figure~\ref{fig:range} shows the uncertainties on the remnant properties for a wider set of BH binary populations.
In particular, we consider three cases with fixed values $q = 0.6$ and $\chi_1 = \chi_2 = 0.3$, $q = 0.6$ and $\chi_1 = \chi_2 = 0.8$,  $q = 0.95$ and $\chi_1 = \chi_2 = 0.8$, as well as a fourth scenario where we distribute the mass ratio according to $p(q) \propto q^{6.7}$ in $[0.25, 1]$ and the spin magnitudes uniformly in $[0.1, 0.8]$. The latter is motivated by current GW detections which favour equal-mass systems \cite{2019ApJ...882L..24A} (but see Ref.~\cite{2020arXiv200408342T}). In all four cases, we consider  both isotropically distributed spins (top panels) and partially aligned spins (bottom panels) at past time infinity, which correspond to distributing $\cos\theta_{i\infty}$ either uniformly in $[-1,1]$ or with a truncated normal distribution centered on 1 with dispersion 0.1, respectively.  We extract 1000 samples from each of these populations. These asymptotic configurations are then evolved 1000 times, sampling the precessional and orbital phases, to obtain corresponding distributions of the remnant properties. We then construct the estimators $\delta M_{\rm f}$, $\delta\chi_{\rm f}$ and $\delta v_{\rm f}$  
of Eq.~(\ref{estimator}) which are shown in Fig.~\ref{fig:range}.

Uncertainties  on $M_{\rm f}$ and $\chi_{\rm f}$ tend to be  smaller for populations where a large fraction of the binaries possess either (i) mass ratio $q$ close to unity, (ii) small magnitudes $\chi_i$, or (iii) small spin misalignments $\theta_{i \infty}$. Binaries with $q\to 1$ result in more similar remnants because the spin magnitude $S$ asymptotes to a constant \cite{2017CQGra..34f4004G}. Systems with either $q\to 0$, $\chi_i\to 0$, or $\theta_{i \infty}\to 0$ are all well described by a single-spin approximation \cite{2008PhRvD..78d4021R,2015PhRvD..91b4043S} where $S$ is also approximately constant. Precession effects are less important for the distribution of $v_{\rm f}$ because the contribution of the orbital-phase sampling is more dominant.  In general, for the populations studied here we find $\delta M_{\rm f}\lesssim 0.004$, $\delta \chi_{\rm f} \lesssim 0.03$, and $\delta v_{\rm f} \lesssim 1.5$.

\subsection{Relation to the spin morphologies}
\label{subsec:morphologies}

We now exploit our evolutionary procedure to highlight correlations between the remnant properties and the precessional dynamics at small separations. We make use of the concept of spin morphology~\cite{2015PhRvD..92f4016G,2015PhRvL.114h1103K,2017CQGra..34f4004G,2019CQGra..36j5003G,2019PhRvD.100l4008P}. The dynamics of BH binaries can be unambiguously classified into three different classes according to the qualitative behavior of the spin angle $\Delta\Phi$ [Eq.~(\ref{angledef})]. In particular, $\Delta\Phi$ can either librate about $0$ and never reach $\pm\pi$ (L0), circulate through the full range $[-\pi,\pi]$ (C), or librate about $\pm\pi$ and never reach $0$ (L$\pi$).
All binaries belong to the circulating morphology at large separation, but GW radiation reaction will generically cause transitions towards the two librating classes before merger.
In the following, we correlate the properties of the merger remnant to the spin morphology evaluated at $\omega_{\rm orb \to sur}$, i.e. right before initializing the NR surrogate.

\begin{figure}[t]
	\centering
	\includegraphics[width=\textwidth]{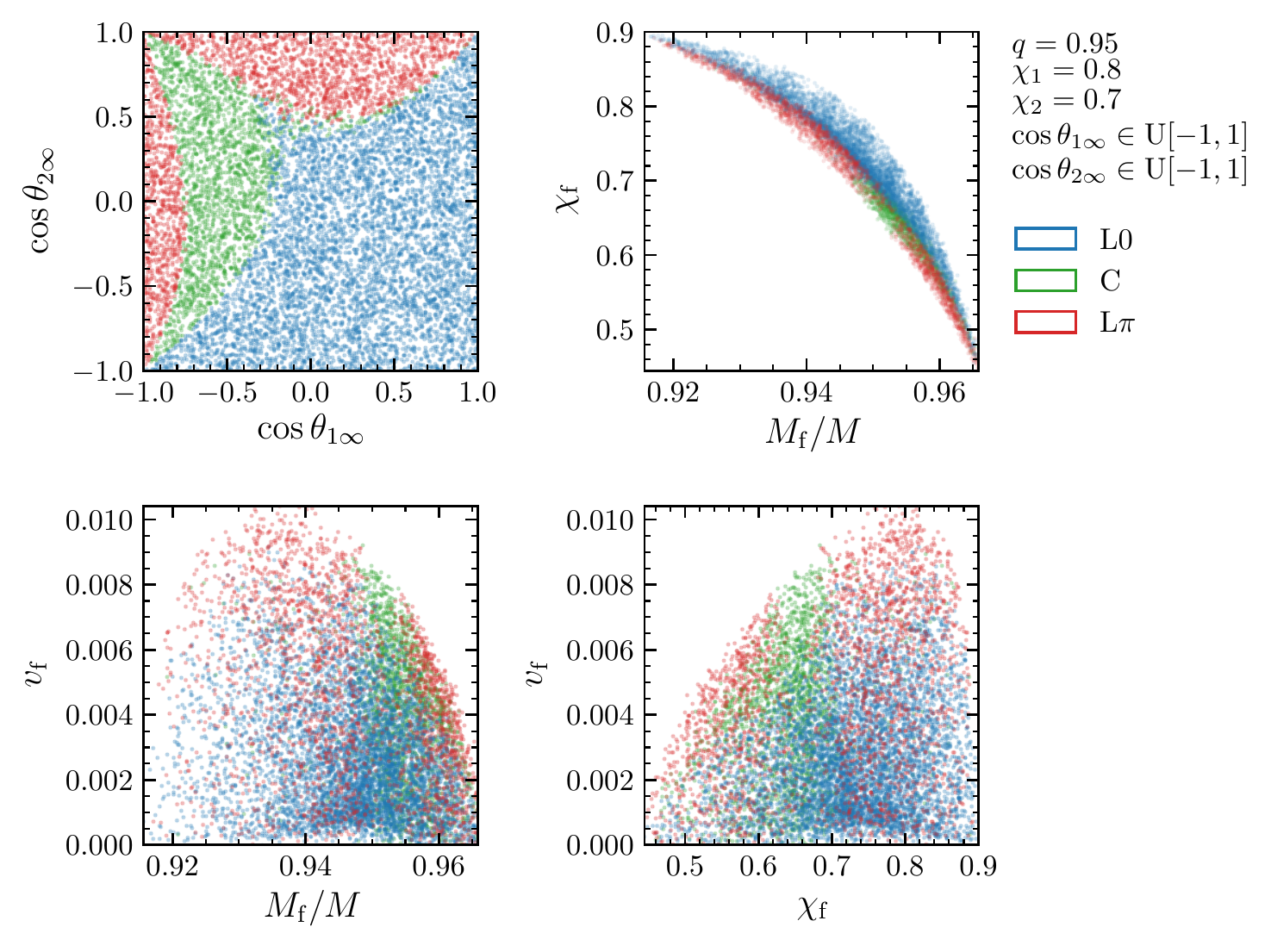}
	\caption{Connecting the asymptotic (top left) and remnant (top right, bottom left, bottom right) parameter spaces using the spin morphologies. We consider a sample of $10^5$ binaries with $q=0.95$, $\chi_1=0.8$, $\chi_2=0.7$ and isotropic spin orientations at $r \to \infty$. Binaries are are coloured according to their spin morphology at $\omega_{\rm orb\to sur}$: blue for binaries librating about $\Delta\Phi=0$ (L0), green for binaries circulating in $\Delta\Phi\in[0,\pi]$ (C) and red for binaries librating about $\Delta\Phi=\pi$ (L$\pi$).}
	\label{fig:scattmorphologies}
\end{figure}

The spin morphology before merger provides a useful mapping between the two asymptotic configurations of the evolution -- the large-separation limit and the post-merger remnant. In Fig.~\ref{fig:scattmorphologies} we illustrate this point showing a population of $10^5$ sources with $q=0.95$, $\chi_1=0.8$,  $\chi_2=0.7$, and isotropic spin directions at large separations. Binaries in each each of the three morphologies tend to cluster in well defined regions of both the ($\theta_{1\infty},\theta_{2\infty}$) and the remnant parameter spaces.\footnote{For fixed $q$, $\chi_1$, $\chi_2$, and $r$, the morphology separation in the ($\theta_{1\infty},\theta_{2\infty}$) plane is exact \cite{2015PhRvD..92f4016G}. In this case, we are evaluating the morphology at fixed orbital frequency $\omega_{\rm orb \to sur}$ which causes minor overlaps between the different classes.}
If $\Delta\Phi$ librates about $0$ ($\pi$), in-plane spin components are closer to alignment (anti-alignment) with each other and produce, on average, final BHs with
masses and spins which are above (below) the hang-up anti-correlation (top-right panel of Fig.~\ref{fig:scattmorphologies})

\begin{figure}[t]
    \centering
    \includegraphics[width=\columnwidth]{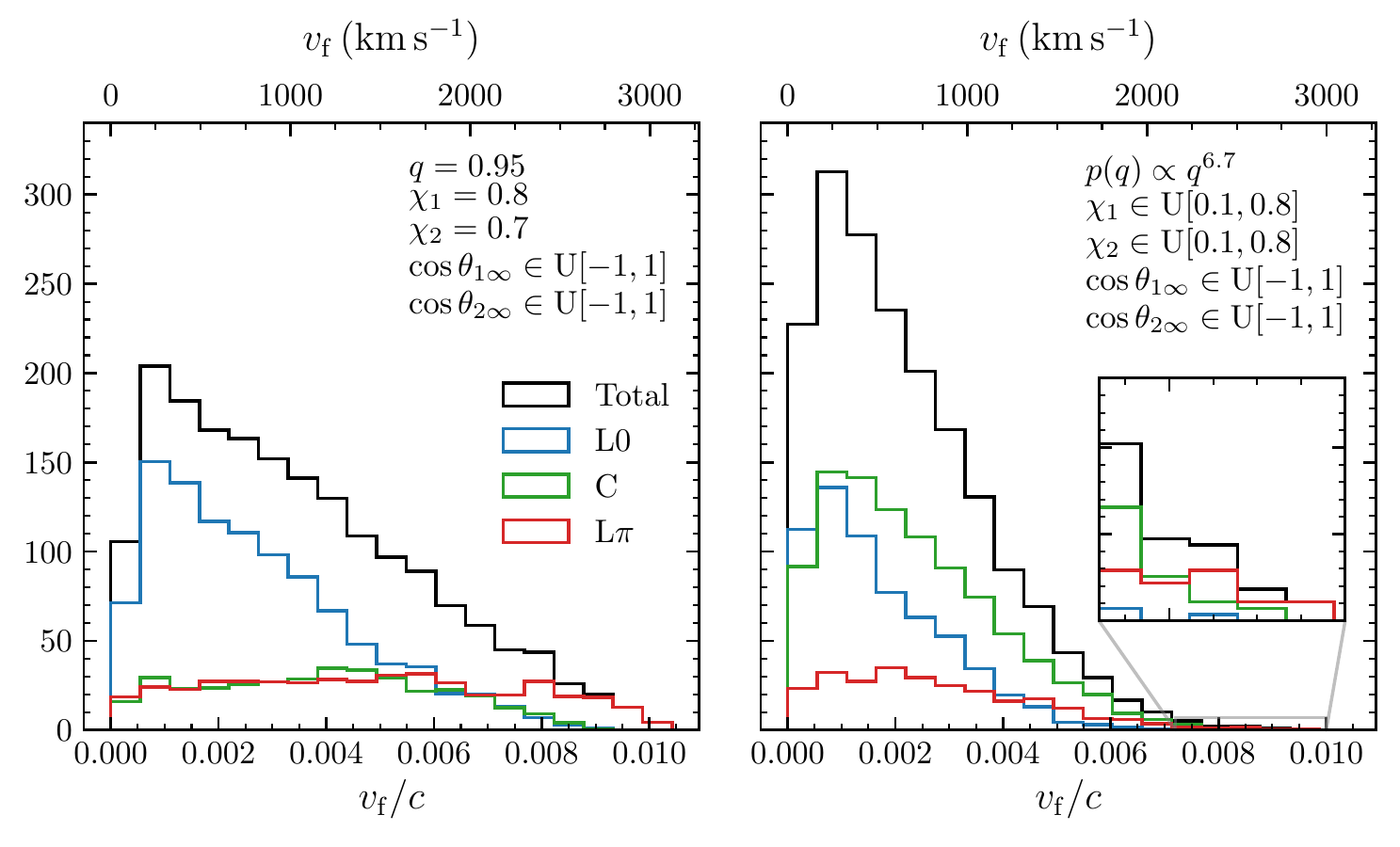}
    \caption{Distributions of the recoil velocities for two binary populations, divided according to the spin morphology at $\omega_{\rm orb \to sur}$. Both samples consist of $10^5$ binaries with isotropic spin directions at $r \to \infty$. The left panel shows a population with fixed values $q=0.95$, $\chi_1=0.8$ and $\chi_2=0.7$. In the right panel we sample mass ratios from $p(q)\propto q^{6.7}$ \cite{2019ApJ...882L..24A} in $[0.25,1]$ and spin magnitudes uniformly in $[0.1,0.8]$. Black histograms correspond to the entire populations; colored histograms indicate the subsets of binaries in either the librating about $\Delta\Phi=0$ (blue), circulating (green), or librating about $\Delta\Phi=\pi$ (red) spin morphology.}
    \label{fig:histokick}
\end{figure}

In the left panel of Fig~\ref{fig:histokick}, we consider the same population and focus on the distributions of the recoil velocities. We find that $\ssim 35$\% ($\ssim 100$\%) of the binaries with $v_{\rm f} \geq 1000$ km/s (3000 km/s) belong to the $\Delta\Phi\sim\pi$ morphology before merger.  This can be also be seen in bottom panels of Fig.~\ref{fig:scattmorphologies}, where red points preferentially populate the regions with large values of $v_{\rm f}$. The role of the spin morphology can be traced back to the very first ``superkick'' NR simulations~\cite{2007PhRvL..98w1101G,2007PhRvL..98w1102C} where BHs were initialized with  $|\Delta\Phi |= \pi$ 
and $\theta_i=	\pi/2$. In this setup, frame dragging of the the two BHs can act constructively, boosting the recoil \cite{2008PhRvD..77l4047B}. Binaries with $|\Delta\Phi|= \pi$ and partial alignment  $\theta_i\lesssim \pi/2$ result in even larger recoils because more energy is emitted~\cite{2011PhRvL.107w1102L}. Conversely, binaries which librate about $\Delta\Phi=0$ before merger present lower kicks of $\mathcal{O}(100)$ km/s and are responsible for the bulk of the population. Circulating sources present a rather flat distribution and allow for most values of $v_{\rm f}$ but the largest ones.

In the right panel  of Fig~\ref{fig:histokick}, we repeat the same exercise but assume a mass ratio distribution $p(q)\propto q^{6.7}$ \cite{2019ApJ...882L..24A} with  $q\in[0.25,1]$ and uniform spin magnitudes $\chi_i\in[0.1,0.8]$. Compared to the previous case, BHs have, on average, lower spins which implies weaker recoils~\cite{2012PhRvD..85h4015L} and a larger fraction of circulating binaries~\cite{2015PhRvD..92f4016G}. The general trends we highlighted above remain valid: the largest (smallest) kicks are largely due to binaries which were librating about $\Delta\Phi=\pi$ ($\Delta\Phi=0$) before merger. 

\subsection{Transition thresholds}
\label{transthr}

\begin{figure}[t]
\centering
\includegraphics[width=\columnwidth]{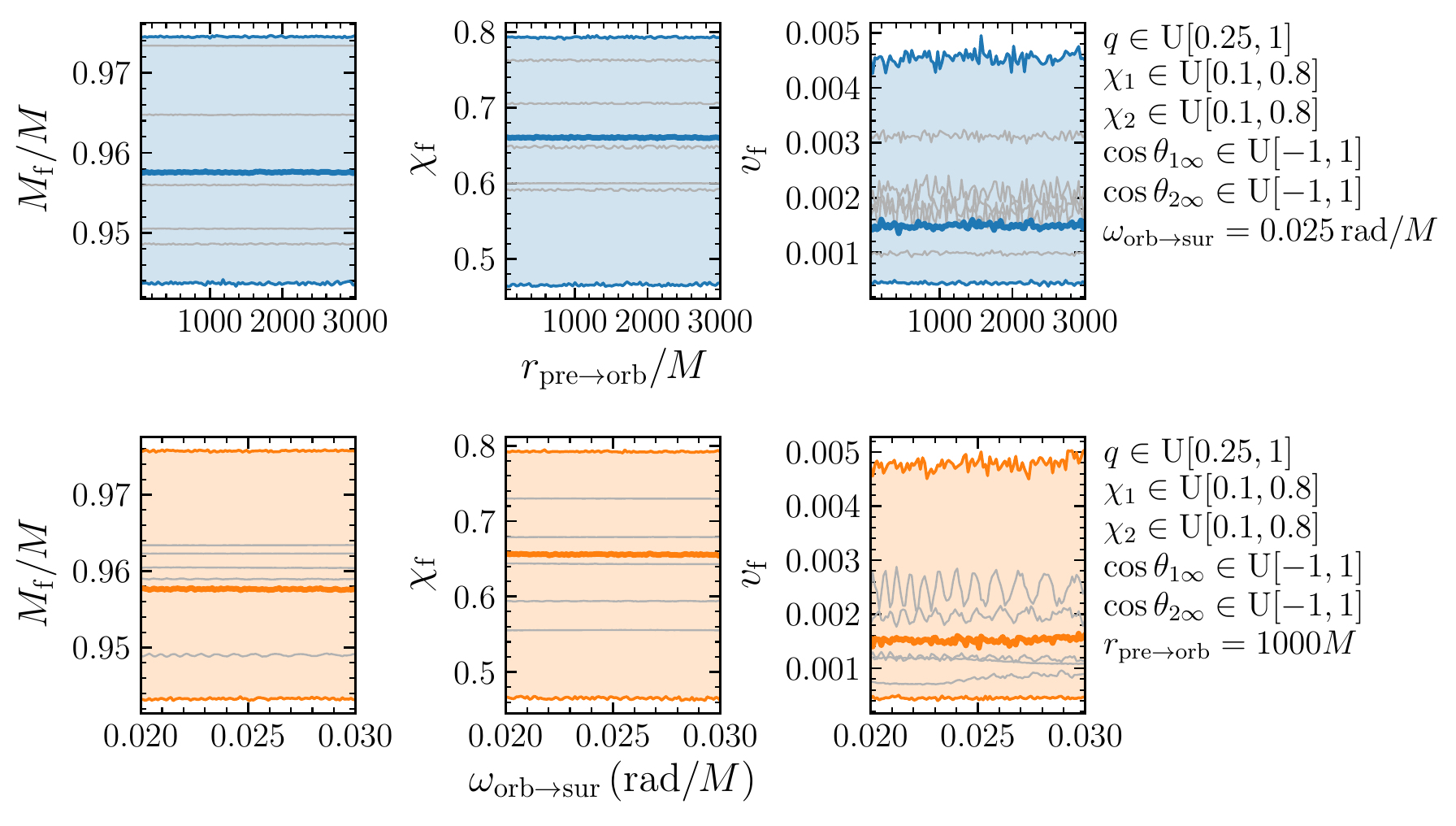}
\caption{Dependence of the 90\% intervals on the transition thresholds $r_\mathrm{pre \to orb}$ and $\omega_\mathrm{orb \to sur}$ for a population of 1000 binaries with $q$ uniformly distributed in $[0.25,1]$, $\chi_i$ uniformly distributed in $[0.1,0.8]$ and isotropic spin directions $\theta_{i\infty}$. In the top row we vary the transition threshold $r_\mathrm{pre \to orb}$ for 100 values spread evenly in $[50M, 3000M]$ and keep $\omega_\mathrm{orb \to sur} = 0.025 \,\mathrm{rad}/M$ fixed. In the bottom row we vary $\omega_\mathrm{orb \to sur}$ for 100 values spread evenly in $[0.02 \,\mathrm{rad}/M, 0.03 \,\mathrm{rad}/M]$ and keep $r_\mathrm{pre \to orb} = 1000M$ fixed. The bottom, middle, and top colored lines in each panel are respectively the $5\%$, $50\%$, and $95\%$ levels of the remnant mass $M_\mathrm{f}$ (left), spin $\chi_\mathrm{f}$ (middle) and kick $v_\mathrm{f}$ (right). In each row we also select five random sources from the population and plot their median levels in gray.}
\label{fig:switch_test}
\end{figure}

Finally, we investigate the robustness of our approach with respect to the two transition thresholds $r_{\rm pre\to orb}$ and $\omega_{\rm orb\to sur}$. Figure~\ref{fig:switch_test} shows how median and $90\%$ interval of the three remnant quantities vary for a population of 1000 binaries with $q$ uniformly distributed in $[0.25,1]$, $\chi_i$ uniformly distributed in $[0.1,0.8]$, and isotropic spin directions $\theta_{\rm i \infty}$. We evolve each source 1000 times, sampling over $S_\mathrm{pre \to orb}$, $\varphi_\mathrm{orb \to sur}$ and the surrogate Gaussian errors, and report the 90\% interval of the remnant properties (Fig.~\ref{fig:switch_test}). We do this for 100 values of $r_{\rm pre\to orb}$ spread evenly in $[50M,3000M]$ with fixed $\omega_\mathrm{orb \to sur} = 0.025\,\mathrm{rad}/M$, and for 100 values of $\omega_\mathrm{orb \to sur}$ spread evenly in $[0.02\,\mathrm{rad}/M,0.03\,\mathrm{rad}/M]$ with fixed $r_{\rm pre\to orb} = 1000M$. The remnant population does not depend sensibly on either of the thresholds in these ranges, though we observe greater discrepancies in $v_\mathrm{f}$ than the other remnant properties. 

We also show five random sources from the same population (gray curves in Fig.~\ref{fig:switch_test}).
The median levels of the remnant mass and spin do not depend strongly on the threshold values, though greater cyclic variation is present in the median levels of the remnant kick velocity. Over the full population this variation is not significant compared to the reported dispersion.

\section{Conclusions}
\label{sec:conclusions}

We presented an evolutionary procedure aimed at capturing the entire history of
coalescing BH binaries. We combined multiple layers of PN integrations
\cite{2015PhRvL.114h1103K,2015PhRvD..92f4016G} and surrogate models trained on
NR simulations \cite{2019PhRvR...1c3015V}. Our scheme is 
robust with respect to the choice of the transition thresholds between the
various techniques we employ~(Sec.~\ref{transthr}).

We map asymptotic binary configurations at $r/M\to\infty$ to the properties of
the BH merger remnant: final mass $M_{\rm f}$, final spin magnitude $\chi_{\rm
f}$, and recoil velocity $v_{\rm f }$. Crucially, this mapping depends on the
sampling of two quantities: the precessional phase $S_{\rm pre \to orb}$ and
the orbital phase $\varphi_{\rm orb \to sur}$. Physically, this is because the
binary completes infinitely many orbits and precession cycles before merger.
Orbital and precessional phases impose a fundamental limit
on the accuracy of models connecting the properties of  astrophysical BH
binary progenitors to the parameters of their merger remnants.

We first concentrated on individual sources and exploited our procedure to assess the impact of these uncertainties on the estimates of the remnant properties. Any given configuration at $r/M\to\infty$ is mapped into a discrete region in the remnant parameter space (see Fig.~\ref{fig:cornerplot}). The relative errors that can be inferred from our results are $\delta M_{\rm f}\sim 0.1\%$ for the remnant masses and $\delta \chi_{\rm f}\sim 1\%$ for the remnant spins. The orbital phase has a major impact on the kick velocity, resulting in $\delta v_{\rm f} \sim\mathcal{O}(1)$. 

We then disentagled the effects of the three different contributions to our uncertainties: the sampling of $S_{\rm pre \to orb}$, $\varphi_{\rm orb \to sur}$, and the NR surrogate errors. The orbital phase plays a dominant role and traces suggestive loop structures in the remnant parameter space (Fig.~\ref{fig:singleloop}). The precessional phase determines the location and the precise shape of these loops.
The remnant uncertainties are found to correlate with the projected effective spin $\xi$ of the binary, with smaller (larger) values of $|\xi|$ presenting large (smaller) dispersion. Uncertainties are more prominent for sources far from the equal-mass and single-spin limits.

Finally, we employed our evolutionary technique to study how the remnant properties correlate with the precessional dynamics at small separations. Merger remnants carry robust information about the precessional spin morphology in the late inspiral. Remnants originating from binaries with different morphological phases at small separations tend to cluster into different regions of the parameter space (Fig.~\ref{fig:scattmorphologies}). Final masses and final spins are strongly anti-correlated: binaries with spins close to (anti-)alignment produce remnants with lower (larger) masses and larger (lower) spins.  BH binaries in the $\Delta\Phi\ssim 0$  ($\Delta\Phi\ssim \pi$) librating morphology result in merger remnants that are systematically above (below) the $M_\mathrm{f}$--$\chi_\mathrm{f}$ correlation. Kicks are also strongly affected, with the vast majority of large recoils originating from binaries in the $\Delta\Phi\ssim \pi$ morphology.

Measurements of BH remnant properties from GW observations~\cite{2019PhRvX...9c1040A,2020PhRvL.124j1104V,2016PhRvD..94b1101G,2016PhRvL.117a1101G} are believed to encode precious (astro)physical information  for stellar-origin (e.g.~\cite{2019MNRAS.482.2991A}), supermassive (e.g.~\cite{2008ApJ...684..822B}), and primordial (e.g.~\cite{2020JCAP...04..052D}) BHs. Our analysis highlights potential caveats and provides a stepping stone to take those ideas to completion.

\section*{Acknowledgments}

We thank Michael Kesden, Ulrich Sperhake, and Emanuele Berti for discussions. 
L.R. is supported by the ERC H2020 project HPC-EUROPA3 (INFRAIA-2016-1-730897).  D.G. is supported by Leverhulme Trust Grant No. RPG-2019-350. Computational work was performed at the Edinburgh Parallel Computing Centre (EPCC), the University of Birmingham BlueBEAR cluster, the Athena cluster at HPC Midlands+ funded by EPSRC Grant No. EP/P020232/1, and the Maryland Advanced Research Computing Center (MARCC).

\section*{ORCID iDs}
Luca Reali~~\orcid{0000-0002-8143-6767}  \href{https://orcid.org/0000-0002-8143-6767}{https://orcid.org/0000-0002-8143-6767}\\
Matthew Mould~~\orcid{0000-0001-5460-2910} \href{https://orcid.org/0000-0001-5460-2910}{https://orcid.org/0000-0001-5460-2910} \\
Davide Gerosa~~\orcid{0000-0002-0933-3579} \href{https://orcid.org/0000-0002-0933-3579}{https://orcid.org/0000-0002-0933-3579} \\
Vijay Varma~~\orcid{0000-0002-9994-1761}  \href{https://orcid.org/0000-0002-9994-1761}{https://orcid.org/0000-0002-9994-1761}

{\small
\setlength{\bibsep}{0.005cm}
\bibliographystyle{iopart-num_leo}
\bibliography{remnantprec}
}

\end{document}